\documentclass[fleqn,10pt]{wlscirep}
\usepackage[utf8]{inputenc}
\usepackage[T1]{fontenc}
\usepackage{xcolor}
\usepackage{enumitem}
\usepackage{mathtools}

\def\bs{\begin{subequations}}
\def\es{\end{subequations}}
\def\aa{\begin{align}}
\def\ab{\end{align}}
\def\ba{\begin{eqnarray}}
\def\ea{\end{eqnarray}}
\def\be{\begin{equation}}
\def\ee{\end{equation}}

\DeclarePairedDelimiter\floor{\lfloor}{\rfloor}

\def\ben{\begin{enumerate}}
\def\een{\end{enumerate}}
\def\bs{\bigskip}

\def\ss{\smallskip}

\title{Contours information and the perception of various visual illusions}

\author[1,2]{Shu Tian Eu}
\author[1,*]{Ee Hou Yong}
\affil[1]{ Division of Physics and Applied Physics, School of Physical and Mathematical Sciences, Nanyang Technological University, Singapore}
\affil[2]{ Department of Physics, University of Wisconsin-Madison, Madison, WI 53706, USA}

\affil[*]{eehou@ntu.edu.sg}

\begin{abstract}
The simplicity principle states that the human visual system prefers the simplest interpretation. However, conventional coding models could not resolve the incompatibility between predictions from the global minimum principle and the local minimum principle. By quantitatively evaluating the total information content of all possible visual interpretations, we show that the perceived pattern is always the one with the simplest local completion as well as the least total surprisal globally, thus solving this apparent conundrum. Our proposed framework consist of (1) the information content of visual contours, (2) direction of visual contour, and (3) the von Mises distribution governing human visual expectation. We used it to explain the perception of prominent visual illusions such as Kanizsa triangle, Ehrenstein cross, and Rubin's vase. This provides new insight into the celebrated simplicity principle and could serve as a fundamental explanation of the perception of illusory boundaries and the bi-stability of perceptual grouping.
\end{abstract}
\begin{document}

\flushbottom
\maketitle
%
%
\thispagestyle{empty}


\section*{Introduction}

A fundamental issue in human perception research is how human subjects show a clear preference for a specific interpretation of visual stimulus from many different ways in which a stimulus could be possibly interpreted. In particular, this phenomenon could be clearly observed when we examine various cases of visual illusions (eg. the Kanizsa illusion which consists of an illusory triangle with three occluded circles instead of three Pac-man inducers) \cite{kanizsa1976subjective}. This gives rise to several interesting questions: does human perception choose an interpretation which is more probable or which is simpler? Are these two interpretations actually equivalent and how do we define the notion of the most probable interpretation and the simplest interpretation? 
  
Over the last century two different paradigms were proposed to explain the human perception of visual stimuli: the likelihood principle\cite{hatfield1985status, von2005treatise, vdHelm2000} and the simplicity principle\cite{Hochberg1953AQA}. The likelihood principle states that visual system has the tendency to perceive the most probable (the one with maximum probability) interpretation while the simplicity principle states that visual system has a preference towards the simplest interpretation of a visual stimulus. These two principles seem to be incompatible, and were always regarded as competitors historically \cite{van2000simplicity, boselie1986test, rock1983logic, pomerantz1986theoretical, sutherland1989simplicity, leeuwenberg1988against}. This debate continued until 1996 when Chater \cite{chater1996reconciling} suggested an interesting new point of view that these two principles could actually be reconciled using Kolmogorov complexity theory. He showed that the visual interpretation which objectively is the most likely (likelihood principle) to be correct is in fact the one with the minimum length of description (simplicity principle). Under this perspective, the most appealing aspect of both principles, namely, the veridicality of perception in terms of external world and the efficiency of the visual system in terms of internal resources are preserved \cite{van2011bayesian}. In recent years the Bayesian model has become a popular choice in unifying these two principles \cite{feldman2009bayes, feldman2013, feldman2014bayesian, gershman2013bayesian, vdHelm2017}. For example, Feldman has shown that the most likely visual interpretation should be the one which is lowest in the partial order of the hierarchical interpretation space (simplicity principle) which maximizes the Bayesian posterior probability (likelihood principle).  
  
In this paper, we aim to address another important question within the framework of simplicity paradigm, which is the contradiction between the global minimum principle and the local minimum principle. The simplicity principle is derived from the well known law of Pragn{\"a}nz.  The law of Pragn{\"a}nz is the most general Gestalt rule that states that people will perceive and interpret ambiguous or complex images as the simplest form(s) possible \cite{wagemans2012century}. This implies that the visual system, like any other physical system, will respond to the stimuli with a tendency to evolve into the equilibrium state involving minimum energy loading. Hochberg and McAllister claimed that in the case of vision, this energy load is, in fact, the information load, and hence proposed that the less the amount of information needed to define a given interpretation of pattern as compared to other alternatives, the more likely that interpretation is perceived \cite{Hochberg1953AQA}. In their original paper, information is defined as the number of different items (eg. number of different line segments, number of different angles) to be described in order to preserve the ``figural goodness", which is a rather vague definition without much mathematical rigour. The global minimum principle was implemented in several perceptual coding languages \cite{leeuwenberg1969quantitative, leeuwenberg1971perceptual, simon1963human, buffart1981coding}, in which it was rephrased and referred to as the minimum principle or the principle of descriptive economy \cite{boselie1989minimum}. In this context the principle states that the perceptually preferred interpretation is the one that requires the fewest predicates in coding models.  

\begin{figure}[!h]
\centering
\includegraphics[width=0.9\textwidth]{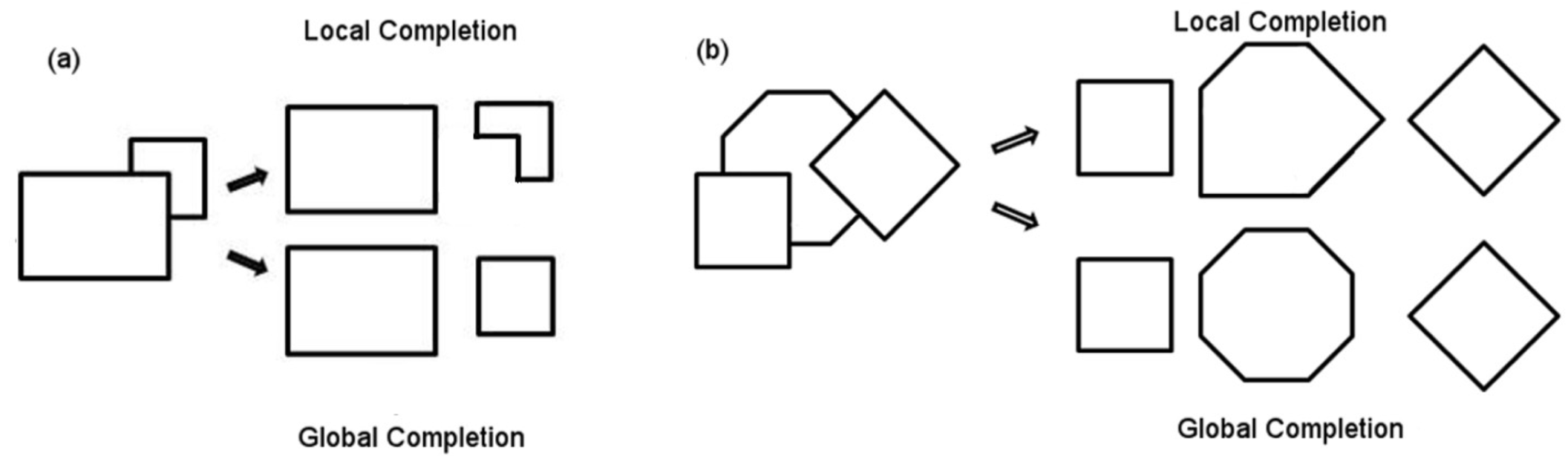}
\caption{{\bf Global completion and local completion of patterns. } (a) An example taken from Boselie, 1988\cite{boselie1988local}. The interpretation predicted by global completion is preferred. (b) Example proposed by Kanizsa to show that the local completion pattern is more prevalent, thus disproving the global minimum principle \cite{kanizsa1985seeing}.  }
\label{pattern}
\end{figure}

However, the global minimum principle was not supported by some theorists \cite{hochberg2007big, rock1983logic}. Kanizsa, for example, asserted that it is the role of good continuation that governs perceptual organization \cite{10009700665, kanizsa1986doppeldeutigkeiten, kanizsa1985seeing}. He believed that the minimum principle should only be applied to local regions of the figure. Our perception and interpretation of figure will not be affected by all global regularities and symmetries within the figure predicted by the global minimum principle. Fig~\ref{pattern} shows two clear examples of how patterns would be completed according to global minimum principle and local minimum principle respectively. It is clear the predicted outcomes are incompatible when used with conventional coding models e.g. structural information theory's formal coding model \cite{van1995competing}, Kolmogorov complexity measure, etc.

In this paper, we aim to resolve this conundrum using the framework of information content of visual contours. By comparing the total information content, namely the surprisal of all possible interpretations, it is observed that the perceived pattern is always the one with the least surprisal. It should be noted that this evaluation of information content is carried out purely geometrically on the final description of the illusory pattern. Unlike conventional coding model, our method shows that the local completed pattern in fact possesses the least amount of information globally, thus reconciling the two seemingly incongruent principles. Human perception selects the pattern with the least total information content based on information along contours. The vagueness of the simplicity paradigm is hence removed. This result also agrees with previous empirical and experimental studies. 

\begin{figure}[!h]
\centering
\includegraphics[scale=1]{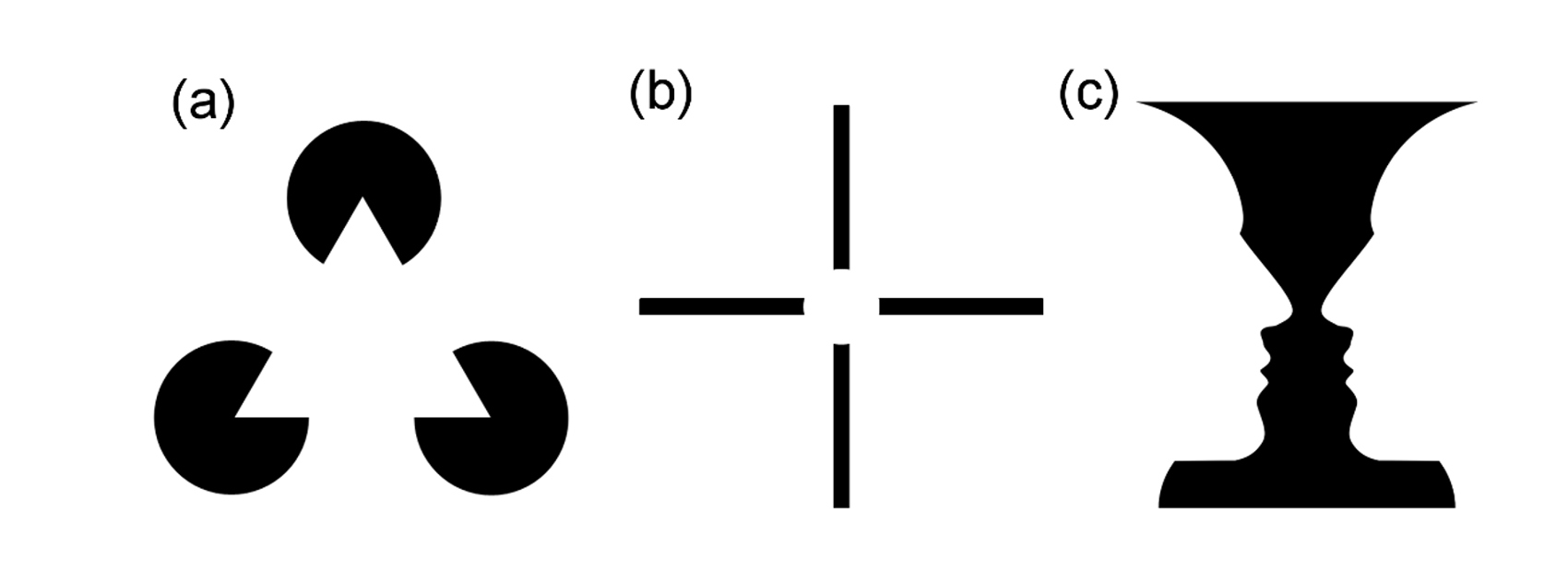}
\caption{ {\bf Three examples of optical illusions.} (a) Kanizsa triangle, (b) Ehrenstein cross and (c) Rubin's Vase.}
\label{illusions}
\end{figure}

This information analysis is also adopted in analyzing three different types of visual illusions (Fig~\ref{illusions}), including Kanizsa illusion, Ehrenstein illusion, \cite{ehrenstein1941abwandlungen} and Rubin's vase \cite{rubin1915synsoplevede}. Kanizsa illusion (Fig~\ref{illusions}a), which has not yet been explained within the framework of global minimum principle, is a class of illusions involving illusory contours, modal and amodal completion \cite{Oliver2016}. Modal completion refers to the process of perceptual completion without occlusion whereas amodal completion refers to the completion of occluded objects \cite{kanizsa1982amodal}. Amodal completion is highly variable since the visual system is free to choose the form of continuity based on local interpretations of the cues provided by the visible part \cite{kanizsa1985seeing, van1995competing}. In the presence of three Pac-man inducers (the circle with a missing piece) on a homogeneous background, we will perceive the figure as a central modal completed triangle that appears brighter and three occluded circles. Although there is no presence of physical boundaries, illusory boundaries will arise along the edges of the triangle. We will show that this illusion arises as perceiving the figure as a triangle and three circles is more economical compared to perceiving it as three circles with missing pieces, as the former contains less information.

Unlike Kanizsa illusions in which illusory contours are induced approximately collinear to the direction of the edges, Ehrenstein illusion is another type of illusion where the illusory contours are induced perpendicularly to the end of the lines (Fig~\ref{illusions}b). It involves the central modal completion of a bright plate and a cross behind the plate. The emergence of the illusory contour will be discussed in detail in a later section. Finally, we discussed the Rubin's Vase, which is a bi-stable two dimensional figure. The Rubin's Vase in Fig~\ref{illusions}c could either be perceived as a central vase or two faces. This bi-stability of figure-ground relation will also be studied within the framework of contours' information.

\section*{Methods}

\begin{figure}[!h]
\centering
\includegraphics[width=0.8\textwidth]{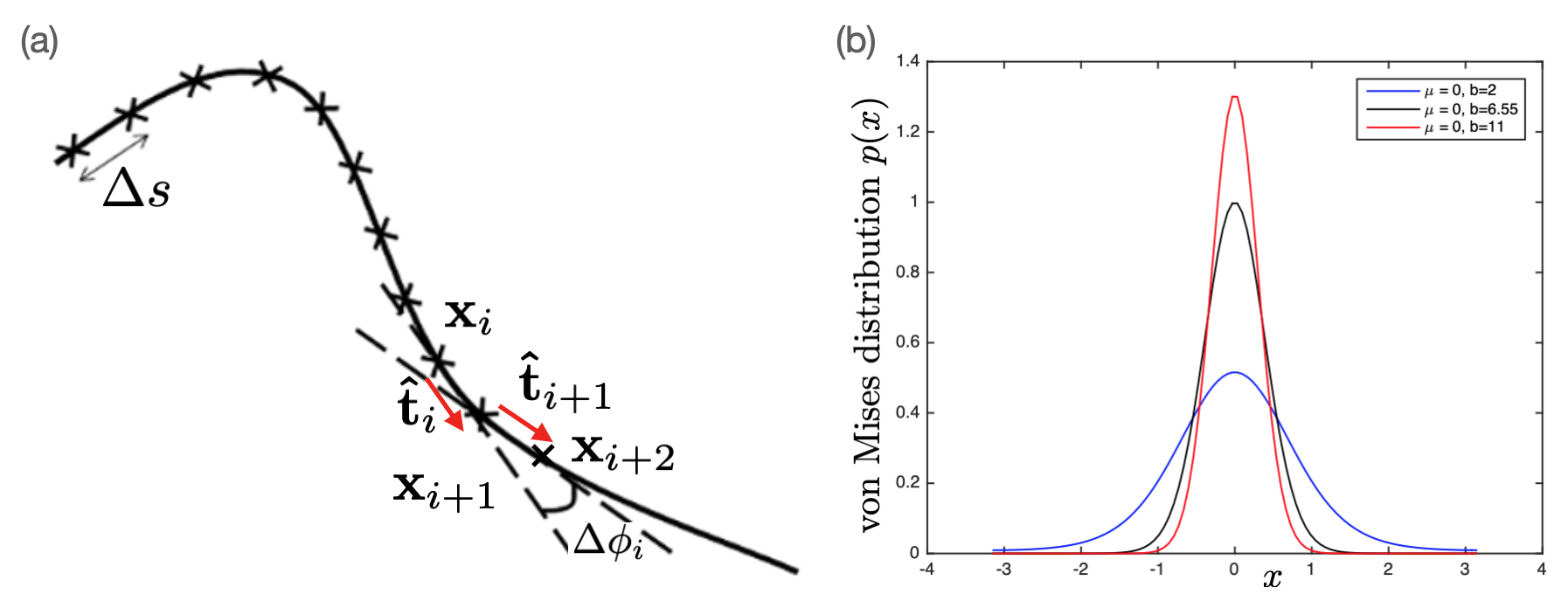}
\caption{{\bf Discrete curve and von Mises distribution.}
(a) A curve that is discretized into $N$ nodes ${\bf x}_i$ that are connected by edges of length $\Delta s$ and tangent ${\bf \hat{t}}_i$. The turning angle between two successive tangent is $\Delta \phi_i$. (b) The von Mises distribution $p(x)$ with different values of $b$ centered at $\mu = 0$ which governs visual expectation on the continuation of smooth curve. The distributions becomes more sharply as the value of $b$ is increased. Note how $p(0) > 1$ if $b > 6.55$.}
\label{framework}
\end{figure}

The concept of surprisal was first introduced by Shannon \cite{BLTJ:BLTJ1338}. In a recent work, Feldman and Singh \cite{feldman} used Shannon's concept of information content and surprisal to quantify contours and object boundaries, a framework that we will adopt. Given a continuous measure $M$ and a probability distribution $p(M)$ that represents our expectation of the value of $M$ before any measurement is taken, what is the information gained by measuring $M$? Shannon claimed that the more the measured value deviates from the expected value, the more ``surprising" it is, and hence the more information it contains. This surprisal, also known as information content, is defined as
\begin{equation}
s(M)=-\log\left[p(M) \right],
\end{equation}
is always positive for a discrete measure and can take on arbitrary real values when the measure $M$ is continuous \cite{Ash1990}. To compute the surprisal along the contours, the targeted curve is sampled at interval of length $\Delta s$, so that a curve of length $L$ will have $N= \floor{\frac{L}{\Delta s}}$ number of straight edges, where$\floor{x}$ is the floor function, e.g. $\floor{3.2} = 3$. The curve is represented by equally spaced $N+1$ vertices ${\bf x}_1, \dots ,{\bf x}_{N+1}$ connected by $N$ straight edges, with unit tangents ${\bf \hat{t}}_1, \dots, {\bf \hat{t}}_{N}$, such that ${\bf \hat{t}}_i = ({\bf x}_{i+1} - {\bf x}_{i})/ ||{\bf x}_{i+1} - {\bf x}_{i}||$ and $||{\bf x}_{i+1} - {\bf x}_{i}|| = \Delta s$ as shown in Fig~\ref{framework}a. For closed curves, ${\bf x}_1 = {\bf x}_{N+1}$. In general, the path length will not be integer multiple of $\Delta s$, i.e. $ N \Delta s \le  L < (N+1) \Delta s$, in which case, we let the last straight edge be of length $||{\bf x}_{N+1} - {\bf x}_{N}|| =  L -  (N-1) \Delta s$. 

The human eye can resolve a periodic signal (e.g. alternating black and white bars) at a spatial frequency of up to about $\theta = 60$ cycles per degree. Assuming that one is viewing the image at a distance of $D = 1$m, this translate to a resolving power of roughly $RP = \theta D = \frac{1}{60} \times \frac{\pi}{180} \approx 3 \times 10^{-4}$m. For all our calculations, we will set the sampling distance to be the human resolution limit i.e.  $\Delta s \approx RP$. Every curve in a visual image, including both closed and illusionary ones, can be broken into a collection of discrete straight edges. From point to point along this sampled curve, the turning angle $\Delta \phi$, which is the angle between two adjacent edges, is measured and serves as the parameter of interest:
\be
\cos (\Delta \phi_i) = {\bf \hat{t}}_{i} \cdot {\bf \hat{t}}_{i+1}.
\ee
Either ${\bf \hat{t}}_{N+1} = {\bf \hat{t}}_{1}$ (for closed curves) or ${\bf \hat{t}}_{N+1}$ may be specified as a boundary condition (for open curves). This choice is motivated by the fact that $\Delta \phi$ is invariant under translational and rotational symmetry, which is the nature of visual perception. Now, having chosen the turning angle $\Delta \phi$ as the continuous measure, we need to choose a probability distribution $p(\Delta \phi)$ that represents our expectation of the value of $\Delta \phi$ before any measurement is taken. In the spirit of simplicity paradigm, it is natural to think that our visual system will expect any curve to continue along its last tangent, since straight line is the simplest extension to any existing pattern. Let us assume that the change in tangent direction on a smooth curve follows the von Mises distribution centered at $\Delta \phi = 0$\cite{feldman} (Fig ~\ref{framework}b) defined on the interval $[-\pi,\pi]$:
\begin{equation}
p(\Delta \phi)=A^\prime(b) \exp{[b\cos{(\Delta \phi)}]},
\end{equation}
where $b$ is the spread parameter that is a measure of the concentration and acts as the inverse of the variance. $A^\prime(b)$ is a normalizing constant independent of $\Delta \phi$ (s.t. $\int_{-\pi}^{\pi} p(\Delta \phi) d\Delta \phi = 1$), and has dependence on $b$ as shown:
\begin{equation}
\begin{split}
A^\prime(b) &= \frac{1}{2\pi I_0(b)},\\
I_0(b)&= \sum_{m=0}^{\infty} \frac{1}{m!\: \Gamma(m+1) }
\left(\frac{b}{2} \right)^{2m}, 
\end{split}
\end{equation}
where $I_0(b)$ is the modified Bessel function of the first kind of order zero. This distribution is chosen as it agrees well with several empirical and experimental studies, including studies conducted by Feldman regarding human subjects' expectation of how a curve is most likely to continue \cite{feldman1997curvilinearity}. Moreover, based on existing research on orientation selectivity of cortical neurons, the tuning curve of the neuronal spike response is best fitted by the von Mises function, which possesses Gaussian-like properties for angular measurement \cite{Hansel4049, 10.1371/journal.pone.0064294, swindale1998orientation}. 


Therefore, for two adjacent points along the contour, ${\bf x}_{i}$ and ${\bf x}_{i+1}$ the information gained, which is the surprisal, is simply:
\begin{equation}
\begin{split}
s(\Delta \phi_i)=-\log{[p(\Delta \phi_i)]}=-\log{(A^\prime(b))}-b\cos{(\Delta \phi_i)} = -\log{(A^\prime(b))} -b \,\, {\bf \hat{t}}_{i} \cdot {\bf \hat{t}}_{i+1}.
\end{split}
\end{equation}
The first term $-\log A^\prime(b)$ has no dependence on the turning angle $\Delta \phi$. The negative cosine dependence on $\Delta \phi$ in the second term shows that the larger the angle deviating from zero (regardless of the direction), the larger the surprisal would be. The total surprisal of a curve, S, is given by
\be
S = \sum_{i=1}^{N} s(\Delta \phi_i) = - \sum_{i=1}^{N} \log{[p(\Delta \phi_i)]}= -b \sum_{i=1}^N {\bf \hat{t}}_{i} \cdot {\bf \hat{t}}_{i+1}  - N \log{(A^\prime(b))}
\ee
Interestingly, this has the same mathematical form as the discretized energy of the wormlike chain model used to describe DNA \cite{Marko95}. For a pattern consisting of $M$ disjoint contours (both real and illusionary), each composed of $N_j$ segments $s_j(\Delta \phi_i)$, where $j = 1, \cdots M$ and $i = 1, \cdots, N_j$, the total surprisal S is
\be
S = \sum_{j=1}^M \sum_{i=1}^{N_j} s_j(\Delta \phi_i) = -  \sum_{j=1}^M \sum_{i=1}^{N_j} \log{[p_j(\Delta \phi_i)]}.
\ee
\section*{Results and Discussions}

\subsection*{Kanizsa Illusions}
\begin{figure}[!h]
\centering
\includegraphics[width=1\textwidth]{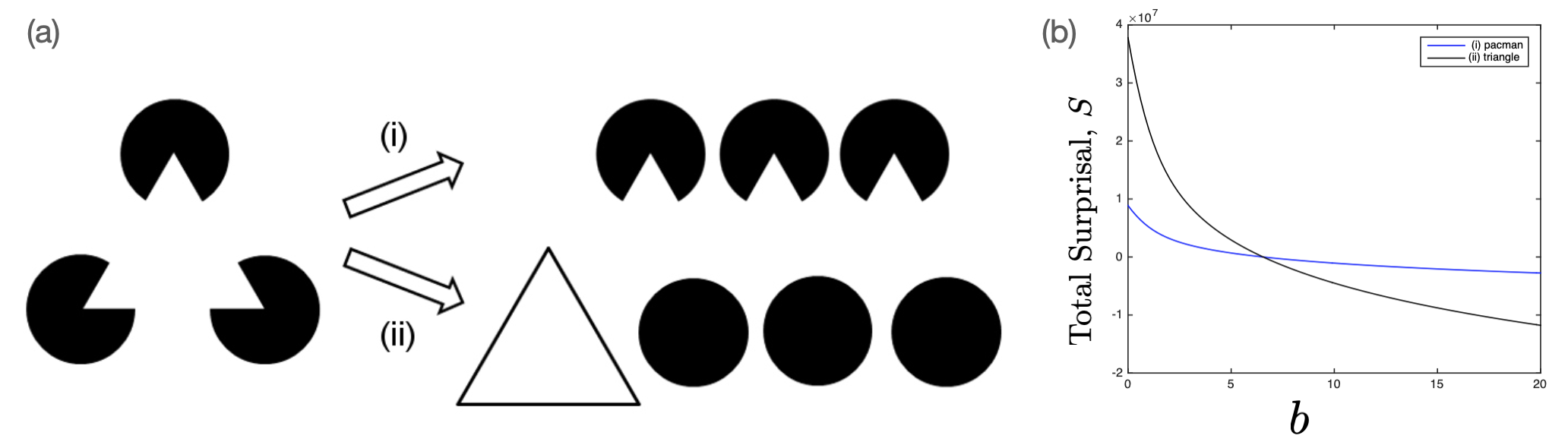}
\caption{{\bf Two ways of interpreting Kanizsa triangles.}
(a) We can either perceive the illusion as (i) three Pac-man inducers or (ii) one modal completed triangle and three amodal completed full circles. (b) Total surprisal as a function of the von Mises parameter $b$ for case (i) (blue curve) and case (ii) (black curve). }
\label{Kanizsa}
\end{figure}

To show that the perception of illusory contours and modal completion are due to the minimum principle of information, we will start with the famous Kanizsa triangles, which has two possible perceptual interpretations. The total surprisal of the two different scenarios: (i) to view the three Pac-man inducers as it is (without modal and amodal completion), and (ii) to view it as a modal completed $\mathcal{N}$-polygon and amodal completed full circles. This is shown in Fig~\ref{Kanizsa}a where $\mathcal{N}=3$ (triangles). If the former case has a higher surprisal than the latter, this shows that our visual system may be conditioned to preferentially select the perception with lower information content. 

\subsubsection*{Kanizsa triangle} 
Kanizsa illusion shows a white colored equilateral triangles with side of length $R$ in the foreground, and three black colored circles of radius $r$ in the background. The distance between the centers of two circles is $R$. For example, the number of edges on a circle of radius $r$ is $N= \floor{\frac{2\pi r}{\Delta s}}$. For typical values $r = 2$cm, $R = 8$cm, $\Delta s = 3 \times 10^{-6}$cm, we find that $N$ is of order $10^6$. The total surprisal for case (i), which consists of three Pac-man inducers is:
\begin{equation}
S^\text{(i)}_\bigtriangleup = -3N \left(\frac{5}{6} + \frac{1}{\pi} \right) \log{(A^\prime(b))} - \frac{5}{2} Nb\cos{\left( \frac{2\pi}{N}\right)}-\frac{3Nb}{\pi}. 
\end{equation}
The total surprisal for case (ii), which consists of three full circles and an illusory triangle is:
\be
 S^\text{(ii)}_\bigtriangleup \approx -3N\log{(A^\prime(b))} \left[1+\frac{R}{2\pi r} \right]-3Nb\cos\left(\frac{2\pi}{N}\right) -  \frac{3NbR}{2\pi r} .
\ee
The total surprisal for this two cases (with the aforementioned parameter values) as a function of the von Mises parameter $b$ is shown in Fig~\ref{Kanizsa}b. By inspection, we see that if $b \ge 6.551$, then $S^\text{(i)}_\bigtriangleup > S^\text{(ii)}_\bigtriangleup$.

Alternatively, we can find the difference in surprisal between the two visual interpretation is $\Delta S_\bigtriangleup =S^\text{(i)}_\bigtriangleup - S^\text{(ii)}_\bigtriangleup$ analytically:
\begin{equation}
\Delta S_\bigtriangleup  \approx 3N \left[\frac{R-2r+ \frac{\pi}{3} r}{2\pi r} \right] \left[\log{(A^\prime(b))} +b \right].
\label{Eq7}
\end{equation}
In order for $\Delta S_\bigtriangleup > 0$, we find that 
\be
\log{(A^\prime(b))}+ b \ge 0.
\label{eq8}
\ee 
This can be further simplified into $b \ge 6.551$ as before. This spread parameter $b$ controls the width of von Mises distribution: the larger the value of $b$, the narrower the distribution (see Fig~\ref{framework}b). At the critical value $b = 6.551$, we see that this is when the maximum of the von Mises distribution at $x=0$ is exactly equal to unity, i.e. $p(0) = 1$. For $b > 6.551$, we see that $p(0) > 1$ and the surprisal $s$ starts to become negative. As long as the von Mises distribution is sharply peaked, outcome (ii) will be preferred.

\subsubsection*{Kanizsa square and polygons}
Following the same treatment, the above results could be generalized to Kanizsa square and even polygons. For the Kanizsa square, there are also two ways of interpreting the stimulus: (i) 4 Pac-man inducers and (ii) a modal completed square with 4 amodal completed circles. The total surprisal difference between this two case is found to be:
\begin{equation}
\Delta S_\square = S^{\text{(i)}}_\square - S^{\text{(ii)}}_\square  \approx 2N \left(\frac{R-2r+\frac{\pi}{4} r}{\pi r} \right) \left[ \log{(A^\prime(b))} +b \right].
\end{equation}
This eventually leads to the same inequality for the spread parameter of the von Mises distribution: $b \ge 6.551$. By induction, it could be proven that the $\mathcal{N}$-sided Kanizsa polygon will have the following total surprisal difference between two different interpretation:
\begin{equation}
\Delta S_{\text{$\mathcal{N}$-polygon}}  =\frac{\mathcal{N}}{2}N \left(\frac{R-2r+\alpha r}{\pi r} \right) \left[ \log{(A^\prime(b))} +b \right],
\end{equation}
where $\alpha = \pi/\mathcal{N}$. This leads to the same inequality $b \ge 6.551$.

\subsection*{Ehrenstein figures}
\begin{figure}[!h]
\centering
\includegraphics[width=0.8\textwidth]{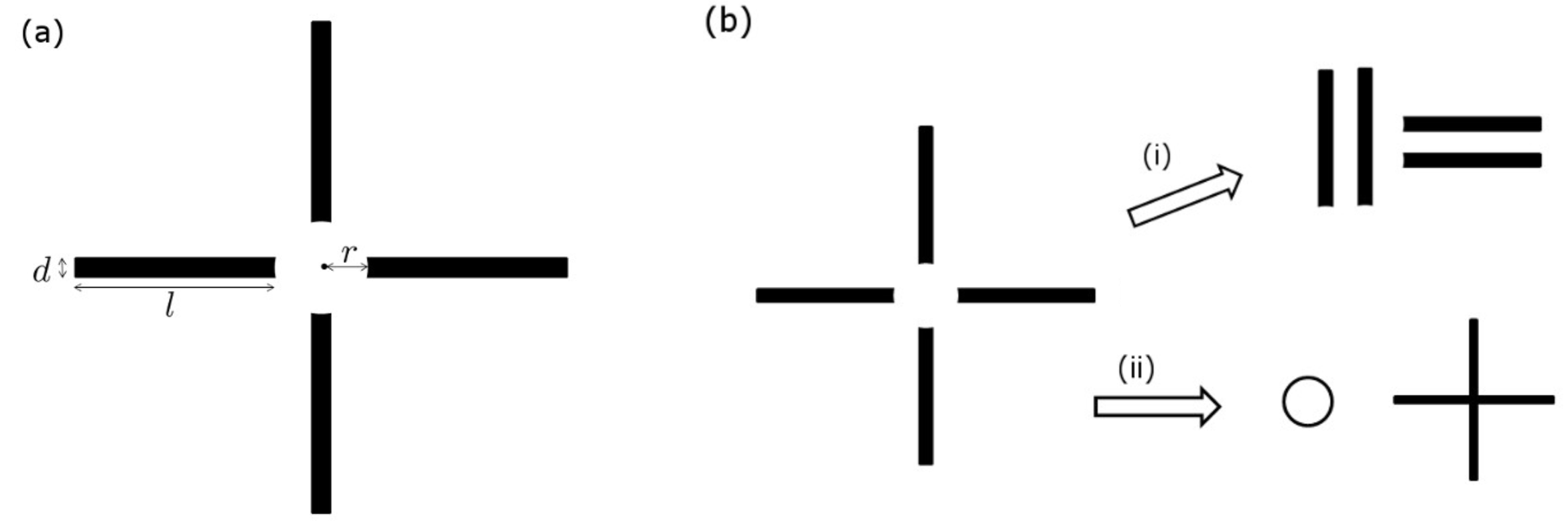}
\caption{{\bf Ehrenstein cross and two ways of interpreting Ehrenstein cross. }{(a) The parameters used for Ehrenstein cross calculation. (b) Two ways to interpret the figure: (i) Four inducing rectangles, (ii) one central modal completed circle with an occluded cross. }
}
\label{Ehrenstein}
\end{figure}

The same analysis can be extended to another type of illusion, Ehrenstein figures, which involves modal completion induced by endpoints. While looking at the Ehrenstein figure, human subjects tend to interpret it as a central bright circle appearing on top of two crossing lines. Again, we will try to explain this preference of interpretation using the surprisal method introduced. Since line width is found experimentally to have positive effect on the clarity of the contour induced \cite{LESHER19932253}, it is natural to consider the inducers as 2D rectangles with certain thickness instead of 1D lines. Fig~\ref{Ehrenstein}a shows 4 rectangular inducers with length $l$ and width $d$ ($l\gg d$) and a central gap of radius $r$. There are two ways to interpret the figure: (i) 4 inducing rectangles and (ii) a central modal completed circle with an occluded cross (Fig~\ref{Ehrenstein}b). The total surprisal difference between case (i) and (ii) is:
\begin{equation}
\Delta S_E=(N+8r-4d-4)[\log{(A^\prime(b))}+b]+4b \approx  N [\log{(A^\prime(b))}+b].
\end{equation}
We see that we arrive at the same inequality as the Kanizsa illusions. As long as $b \ge 6.551$, (ii) is the preferred interpretation to case (i). 

\subsection*{Global minimum and local minimum }
From previous discussion, it is clear that the perception of illusory contours could be explained by comparing the total surprisal of different possible figural interpretations. Now, the difference and advantage of this framework compared to previous coding models shall be discussed. As shown in Fig~\ref{pattern}, global minimum principle and local minimum principle predict different interpretations. For the example shown in Fig~\ref{pattern}(a), the interpretation predicted by global minimum principle is preferred experimentally. On the other hand, for the example shown in Fig~\ref{pattern}(b), the result predicted by local minimum principle is shown to be more prevalent experimentally \cite{kanizsa1985seeing, van1995competing}. This introduces an ambiguity into the selection process of the most plausible interpretation as these two principles predict different outcomes. However, by calculating information content under our framework, it could be proven that the preferred interpretations in both cases in fact possess the lowest surprisal compared to their counterparts. The case in Fig~\ref{pattern}(b) is especially worth noting as we show that the result predicted by local minimum principle in fact possesses lower surprisal than the other, and hence it is the minimum information configuration globally. This sheds a light into combining these two principles, as they could be equivalent under the new way of evaluating the information content of the interpretation of figure. 

\subsection*{Rubin's Vase}
Other than explaining illusions involving illusory contours and resolving the difference between global and local minimum, this method can also be used to explain the figure-ground illusion, which is the bi-stability of perceptual grouping. In our previous discussion, the mean of von Mises distribution is zero, which implies that our perception is insensitive to the propagation direction of tangent, i.e. turning clockwise $\Delta \phi < 0$ or counterclockwise $\Delta \phi >0$ of the tangent yield the same amount of information. Under this situation where we do not have a preferred direction, the expected change of the tangent is perfectly described by a symmetric von Mises distribution centered at zero. It is obvious that in the example of Rubin's vase, viewing the vase or viewing the faces possess the same amount of surprisal, as the boundary contours are shared by these two interpretation, thus causing the ambiguous effect in this kind of bi-stable illusion. 

\begin{figure}[!h]
\centering
\includegraphics[width=1.0\textwidth]{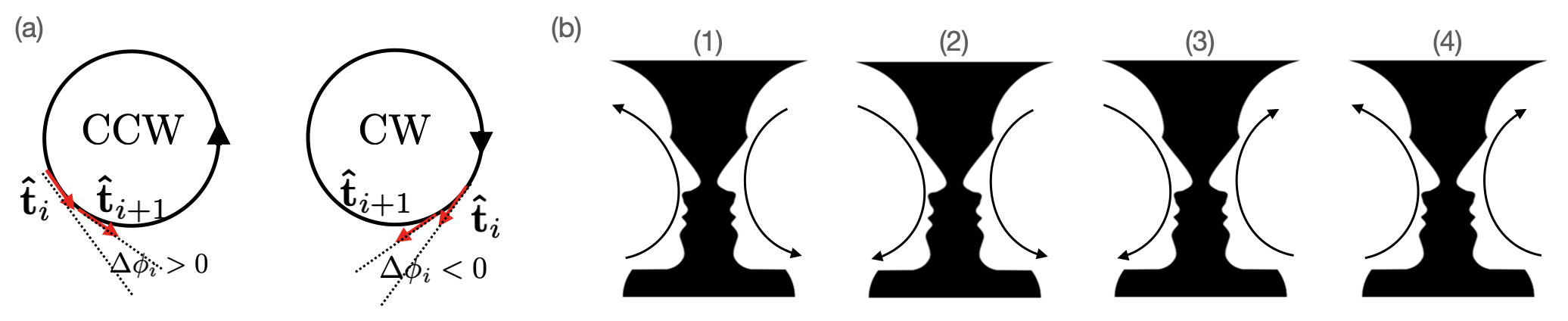}
\caption{{\bf Total surprisal for circle and Rubin's vase.} (a) The total surprisal for circle drawn in clockwise and counterclockwise sense will be different if the underlying von Mises distribution is skewed. (b) Adding directions to the two contour lines in the Rubin's vase image. There are a total of four different combinations, leading to different total surprisal.}
\label{Rubin}
\end{figure}

However, if a viewer is ``biased" so that turning in clockwise (CW)/counterclockwise (CCW) sense is preferred, the von Mises distribution governing the visual expectation is no longer centered at zero, but skewed towards left/right as shown in Fig~\ref{Rubin}a. The resulting surprisal is
\be
p(\Delta \phi)=A^\prime(b) \exp{[b\cos{(\Delta \phi \pm \delta )}]},
\ee
where $\delta > 0$ is the degree of skewness and the plus/minus sign is the preference for CW/CCW rotations \cite{feldman}. Consecutive straight edges, i.e. ${\bf \hat{t}}_{i} = {\bf \hat{t}}_{i+1}$, rather than being the most expected case as before, are now slightly surprising. For the minus case (preference for CCW), this means that consecutive tangents that turn in the CCW sense carry greater information than otherwise equivalent CW tangents and vice versa. This broken asymmetry means that the way we ``draw" the visual contours is now important. Consider a person biased towards CCW rotations with skewness $\delta = \Delta \phi = \floor{\frac{\Delta s}{r}}$. By tracing the circle in a CW manner $(\Delta \phi = -\delta<0)$, the total information content is
\be
S_{\text{CW}} = - N \log \left(A^\prime(b) \exp(b \cos(-|\Delta \phi| - \delta)) \right) = - N \log \left(A^\prime(b) \exp(b \cos(2 \delta)) \right).
\ee 
On the other, when visually tracing in a CCW manner $(\Delta \phi = \delta >0)$, the total information content is 
\be
S_{\text{CCW}} = - N \log \left(A^\prime(b) \exp(b \cos(|\Delta \phi| - \delta)) \right) = S_{\text{CW}} = - N \log \left(A^\prime(b) \exp(b) \right).
\ee
The difference in surprisal is 
\be
\Delta S = S_{\text{CW}} - S_{\text{CCW}} \approx 2Nb \delta^2 > 0.
\ee
Thus we see that the direction of the contour lines will determine the total surprisal for the case of a skewed von Mises distribution. For a general picture with multiple contour lines, the choice of contour directions with the least total information content would be preferred.

Referring back to the Rubin's vase illusion, there are four ways to add directions to the two contour lines, enumerated by (1) to (4) as shown in of Fig~\ref{Rubin}b. A person with a preference for CCW rotations will prefer to follow the contours in the sense shown in (1) since it has the lowest total surprisal. In fact, we find that
\be
S^{\text{CCW}}_{(1)} < S^{\text{CCW}}_{(2)} =  S^{\text{CCW}}_{(4)} < S^{\text{CCW}}_{(3)}.
\ee
The continuity of the contour directions together with the preference for CCW rotations meant that we get two disjoint parts. The directions of the contours is such that the face becomes the ``figure" and the vase becomes the ``ground" and the resulting interpretation is that of two faces. On the other hand, a person who has a predisposition for CW rotations, we get
\be
S^{\text{CW}}_{(3)} < S^{\text{CW}}_{(2)} =  S^{\text{CW}}_{(4)} < S^{\text{CW}}_{(1)}.
\ee
Hence, the contours directions of (3) would be the preferred outcome. In this case, the continuity of the contour directions and the predilection for CW rotations results in one connected big part. In this case, vase is now the ``figure" while the two faces are the ``ground" and the resulting interpretation is that of a vase. Here we see that the formation of ``figure" and ``ground" that results from the choice of contour directions will tilt the optical illusion towards different interpretations.

In summary, in this paper we give a brief review on the status of the long celebrated simplicity principle, which is derived from the law of Pragn{\"a}nz and later referred to as the global minimum principle. We have shown that by employing Feldman and Singh's method in calculating the surprisal along contours and choosing a suitable spread parameter for the von Mises distribution governing human visual expectation, visual illusions involving the perception of illusory contours such as Kanizsa illusions and Ehrenstein figures could be well explained. Unlike conventional coding model in which the contradiction between the global minimum principle and local minimum principle is inevitable, this method naturally resolves the contradiction to some extent by showing that the prediction by the local minimum principle, which is experimentally proven to be more prevalent, in fact possesses the globally minimum surprisal. The bi-stability of perceptual grouping, for example in the case of the Rubin's vase is also studied. Individual with biased visual expectation, in which his or her perception is governed by skewed von Mises distribution, has a tendency towards perceiving one interpretation over another based on the choice of contour directions. 

\bibliography{main}



\section*{Acknowledgements (not compulsory)}

S.T.E. and E.H.Y. acknowledge support from Nanyang Technological University, Singapore, under its Start Up Grant Scheme (04INS000175C230).

\section*{Author contributions statement}
S.T.E. and E.H.Y. conceived the project; S.T.E. and E.H.Y. carried out the calculations; S.T.E. and E.H.Y. wrote the paper.






\end{document}